\documentclass[sigconf,natbib=false,nonacm]{acmart}
\RequirePackage[
  datamodel=acmdatamodel,
  style=acmnumeric,
  ]{biblatex}

\usepackage{listings}
\usepackage{xcolor}

\usepackage{siunitx}
\usepackage{booktabs}
\usepackage{float}
\usepackage{pifont}
\usepackage{todonotes}
\usepackage{multirow}
\newcommand{\cmark}{\ding{51}} 

\usepackage{framed}
\definecolor{shadecolor}{RGB}{245,245,245} 

\newenvironment{finding}{%
  \begin{shaded}\noindent\textbf{Key Takeaways.} }%
  {\end{shaded}}

\begin{document}

\title{An Empirical Comparison of Security and Privacy Characteristics of Android Messaging Apps}
\renewcommand{\shorttitle}{Comparing Privacy Characteristics of Android Messaging Apps}

\newcommand{\na}[1]{\todo[color=blue!20,inline]{#1}}


\author{Ioannis Karyotakis} 
\authornote{These authors contributed equally to this work.}
\affiliation{%
 \institution{AUEB \& NTUA}
 \city{Athens}
 \country{Greece}
}
\email{karyotakisg@aueb.gr}
\orcid{0009-0004-2590-2737}

\author{Foivos Timotheos Proestakis}
\authornotemark[1]
\affiliation{%
 \institution{AUEB \& NTUA}
 \city{Athens}
 \country{Greece}
}
\email{proestakis@aueb.gr}
\orcid{0009-0008-3371-7338}

\author{Evangelos Talos}
\authornotemark[1]
\affiliation{%
 \institution{AUEB \& NTUA}
 \city{Athens}
 \country{Greece}
}
\email{vtalos@aueb.gr}
\orcid{0009-0009-7789-2135}

\author{Diomidis Spinellis}
\affiliation{%
 \institution{AUEB \& TU Delft}
 \country{Greece \& Netherlands}}
 \email{dds@aueb.gr}
 \orcid{0000-0003-4231-1897}

\author{Nikolaos Alexopoulos}
\affiliation{%
\institution{AUEB}
\city{Athens}
\country{Greece}}
\email{alexopoulos@aueb.gr}
\orcid{0000-0001-8383-4761}

\begin{abstract}
Mobile messaging apps are a fundamental communication infrastructure, used by billions of people every day to share information, including sensitive data. Security and Privacy are thus critical concerns for such applications. Although the cryptographic protocols prevalent in messaging apps are generally well studied, other relevant implementation characteristics of such apps, such as their software architecture, permission use, and network-related runtime behavior, have not received enough attention.
In this paper, we present a methodology for comparing implementation characteristics of messaging applications by employing static and dynamic analysis under reproducible scenarios to identify discrepancies with potential security and privacy implications. We apply this methodology to study the Android clients of the Meta Messenger, Signal, and Telegram apps.

Our main findings reveal discrepancies in application complexity, attack surface, and network behavior. Statically, Messenger presents the largest attack surface and the highest number of static analysis warnings, while Telegram requests the most dangerous permissions. In contrast, Signal consistently demonstrates a minimalist design with the fewest dependencies and dangerous permissions. Dynamically, these differences are reflected in network activity; Messenger is by far the most active, exhibiting persistent background communication, whereas Signal is the least active. Furthermore, our analysis shows that all applications properly adhere to the Android permission model, with no evidence of unauthorized data access.
\end{abstract}


\maketitle

\section*{Note}
This publication is an extended version of the ACM poster  ``An Empirical Comparison of Security and Privacy Characteristics of Android Messaging Apps'', presented at the 41st ACM/SIGAPP Symposium on Applied Computing (SAC~'26), March 23--27, 2026, Thessaloniki, Greece \href{https://doi.org/10.1145/3748522.3779858}{10.1145/3748522.3779858}.

\section{Introduction}
Mobile messaging applications have become the dominant medium for remote communication, gradually replacing legacy channels such as SMS and phone calls. Billions of users worldwide rely on them for exchanging often-sensitive personal and professional information. Consequently, the security and privacy (S\&P) properties of these applications are of critical societal relevance. Although the cryptographic protocols underlying message confidentiality, such as the Signal Protocol, have been formally studied~\cite{gordon2017signal, gordon2020signal}, less attention has been directed toward the actual implementation and execution characteristics of messaging applications.

S\&P guarantees of real-world messengers ultimately depend not only on the correctness of their underlying protocol but on a series of factors, including their overall implementation quality and their (meta)data collection and sharing practices. Prior work studying relevant characteristics of other popular apps, e.g. contact tracing~\cite{leith2021contact, leith2023contact}, parental control~\cite{feal2020angel}, pre-installed~\cite{gamba2020preinstalled, liu2023android}, paid vs. free~\cite{han2020price}, or even the Google Dialer and Messenger~\cite{leith2022google}, found that many of them request excessive permissions or collect and transmit data beyond user expectations. While many app categories have been previously studied, messaging apps present a notable gap.

In this paper, we introduce a hybrid methodology for comparing the S\&P characteristics of mobile messaging applications.
Our method combines static and dynamic analysis to answer the central question: \emph{How do messaging apps differ in their S\&P-relevant implementation characteristics and what are the implications of those differences?} We study apps along two complementary dimensions:
\smallskip\par\noindent\emph{-- Static characteristics:} These include metrics related to complexity, attack surface, requested permissions, and privacy-relevant warnings raised by static analysis tools.
\par\noindent\emph{-- Dynamic characteristics:} These include runtime network metrics, such as connections, peers, and bytes sent/received over the network under well-defined reproducible test scenarios, in addition to compliance with permission restrictions.

\smallskip Methods used in previous studies for the dynamic analysis of other application categories are mostly based on network traffic inspection and are not applicable in our setting. Messaging apps make extensive use of custom certificate pinning~\cite{pradeep2022pinning}, making it especially difficult to obtain the plaintext of encrypted network communication, even with specialized tools.
To overcome this challenge, we trace app behaviors at a lower level using Linux kernel traces.\footnote{https://github.com/SliceDroidTeam/SliceDroid} This allows us to track resource access under different permission settings and usage scenarios.

To demonstrate the efficacy of our approach, we perform the first, to the best of our knowledge, empirical study on the S\&P characteristics of Android messaging applications. To strike a balance between breadth and depth, we select three applications for our study: Meta Messenger, Signal, and Telegram. We select these applications based on two criteria. First, they represent a diverse spectrum of design philosophies, business models, and source-code availability. Second, their significant collective popularity, with over 6 billion downloads from the Google Play Store,\footnote{\url{https://signal.org} 100+ million downloads, \url{https://telegram.org} 1+ billion downloads, \url{https://www.messenger.com} 5+ billion downloads, as of October 2025} makes them highly relevant case studies for applying our methodology.
Specifically, Messenger serves as an example of a fully proprietary application integrated into a large, data-driven corporate ecosystem. Signal represents a fully open-source project (both client and server), developed by a non-profit foundation with an explicit focus on maximizing user privacy. Finally, Telegram occupies a middle ground, featuring open-source clients but a proprietary, closed-source server infrastructure, supported by a for-profit model.

\smallskip\par\noindent\textbf{Contributions.}
Our contributions are the following:
\begin{itemize}
\item We introduce a methodology to identify static and runtime discrepancies between messaging app implementations in a reproducible manner, overcoming limitations of previous analysis approaches.
\item We use this methodology to identify discrepancies between three popular messaging apps, performing the first empirical study on their implementation characteristics.
\item We analyze and discuss the identified discrepancies and highlight their practical implications.
\end{itemize}

\smallskip\par\noindent\textbf{Main findings.} \textbf{Signal} has a minimalist implementation, featuring the fewest sensitive permissions, the smallest overall attack surface, and the lowest level of network activity. \textbf{Messenger} shows comparatively high network activity, especially in the background. It also exposes the largest attack surface, exporting nearly twice as many components as Signal and Telegram. \textbf{Telegram}, in contrast, requests the highest number of runtime permissions and receives the highest volume of traffic from the network, especially when permissions are restricted. In general, these discrepancies reflect different implementation choices of the three messaging clients.

\smallskip\par\noindent\textbf{Availability.} We make our code and data available to the community for further studies.\footnote{\url{https://zenodo.org/records/17306016}}

\section{Background} We provide some information on the Android platform and the analysis techniques employed in mobile privacy research.

\smallskip\par\noindent\textbf{The Android Application Model}
Android is a Linux-based OS with over three billion active users\footnote{\url{https://www.businessofapps.com/data/android-statistics/}}, making it the most popular OS in the world. 
Android applications are distributed as Android Packages (APKs). Each app contains an \texttt{AndroidManifest.xml} file, which serves as its blueprint, declaring essential information such as its components, required permissions, and hardware features~\cite{android_manifest_overview}. Applications are built from four primary types of components: Activities, which provide the user interface; Services, which perform long-running operations in the background; Broadcast Receivers, which respond to system-wide announcements; and Content Providers, which manage and share application data. A critical security attribute for these components is the \texttt{android:exported} flag in the manifest. When set to true, a component is made accessible to other applications on the device.

\smallskip\par\noindent\textbf{The Android Permission System} To protect user data and system resources, Android employs a permission-based access control model built on the principle of sandboxing. Each application operates in its own isolated environment~\cite{android_sandbox, enck2011study}. Access to sensitive data (e.g., contacts, location) or system features (e.g., camera, microphone) requires the application to explicitly request the corresponding permission. Permissions are categorized by their \texttt{protectionLevel}~\cite{android_permissions_overview}. Of particular interest are dangerous permissions, which grant access to private user data or control over the device in ways that can negatively impact the user. Since Android 6.0 (API level 23), these permissions must be requested from the user at runtime and can be revoked at any time~\cite{android_permissions_overview}.

\smallskip\par\noindent\textbf{Static and Dynamic Analysis}
Mobile application analysis typically relies on two complementary approaches. 
\emph{Static analysis} inspects an app's source code or application package (APK) without execution to examine its declared properties, such as permissions and exported components, and to identify third-party libraries or potential vulnerabilities in its code.
\emph{Dynamic analysis} monitors an app's runtime behavior by collecting low-level data, such as network packets (\texttt{tcpdump}) and kernel-level system call traces (\texttt{ftrace}). A core challenge in Android is the attribution problem: due to extensive Inter-Process Communication (IPC), system services often act on an app's behalf, obscuring an action's origin. To solve this, dynamic analysis tools like SliceDroid~\cite{Alexopoulos2025SliceDroid} efficiently track data flow by considering IPC. As detailed in Section~\ref{sec:methods}, our work extends this approach to reliably attribute network activity.

\section{Related Work}
Mobile application privacy has been a long-standing concern, and a substantial body of work has examined S\&P characteristics of mobile apps.

An important line of work is based on the capture and analysis of the outbound communication content of applications (dynamic analysis) to identify potential privacy leaks.
A longitudinal analysis of over 500 popular Android applications found that newer app versions increasingly transmit more types of personally identifiable information, while the adoption of transport-layer encryption has been slow~\cite{ren2018bug}. The problem is further exacerbated by pre-installed system components, particularly in some of the largest smartphone markets. A study of devices from popular vendors in China revealed that numerous pre-installed apps are granted dangerous privileges and transmit sensitive data---including persistent identifiers, GPS coordinates, and call history---to third-party domains without user consent or notification~\cite{liu2023android}. Other studies have investigated privacy risks in applications designed for public welfare, such as contact tracing apps~\cite{leith2021contact, leith2023contact}. While apps managed by health authorities were often found to be privacy-respecting, the underlying third-party services, such as Google Play Services, collected extensive telemetry data, including hardware identifiers, and made frequent network connections that could enable IP-based location tracking. A large-scale study by \textcite{reardon201950} uncovered the active use of both side channels (exploiting system flaws to read data) and covert channels (collusion between apps to share data) in hundreds of popular applications. This allowed unauthorized access to sensitive information such as unique device identifiers and geolocation, effectively bypassing Android's permission model. The dynamic analysis aspect of our work differs from all aforementioned approaches, as we found messaging apps to employ difficult-to-circumvent certificate pinning, making the capture of communication content especially challenging. Instead, we trace app operations in the kernel level to identify discrepancies and interesting patterns.

Another line of work performs privacy analysis of mobile apps using static and dynamic analysis tools and manual inspection. An investigation of popular mental health apps~\cite{iwaya2023privacy} revealed insecure cryptography, the leakage of personal data in plaintext, and the extensive use of third-party libraries that facilitate user profiling and tracking. A study on parental control apps~\cite{feal2020angel} identified similar problems, including a lack of transparency, as many apps were found to collect and transmit personal information without user consent. Another large-scale study on differences between paid and free variants of Android apps~\cite{han2020price} found no evidence of better privacy protection for paid versions. To the best of our knowledge, there are no similar studies for messaging apps.

With regard to messaging applications, prior work is largely focused on testing how specific mechanisms could compromise user privacy. 
Systematic forensic examinations of WhatsApp, Telegram, Signal, and Viber revealed that chat databases, cached media, and push-notification logs can expose conversation partners and group identifiers, sometimes persisting despite user deletions~\cite{anglano2014forensic, lone2015implementation, anglano2017forensic, krishnapriya2021forensic}. A recent study by \textcite{kirchner2024black} used ``honey messages'' containing unique URLs to test whether messaging services inspect message content. Their findings revealed that a majority of mobile messaging apps fetch these links, indicating server-side analysis of user messages, at minimum for generating link previews. \textcite{samarin2024medium} analyzed 21 secure messaging apps and found that over half leaked personal data, such as message content or user identifiers, to Google through Firebase Cloud Messaging (FCM) payloads—a transfer rarely disclosed in their privacy policies. Compared to these studies, we have a wider focus on app characteristics and behaviors, not focusing only on specific mechanisms, like push notification.

\section{Methods}
\label{sec:methods}
Our approach resembles differential testing for software~\cite{mckeeman1998differential}.
We use a combination of methods (static/dynamic analysis, network data inspection) to identify \emph{discrepancies} between applications and investigate these discrepancies to assess their S\&P implications.

\subsection{Static analysis}

To ensure reproducibility, official production-signed APKs were extracted from physical Android devices via the Android Debug Bridge (ADB). The retrieved packages were subsequently examined with established tools, as follows:

\par\noindent\textbf{Structural \& Manifest Analysis.}
Initial inspection utilized core utilities from the Android SDK \footnote{\url{https://developer.android.com/studio/command-line}}. The \texttt{aapt} (Android Asset Packaging Tool), along with \texttt{apkanalyzer}, was employed to extract metadata from the \texttt{AndroidManifest.xml} and to dissect the package composition. This process allowed us to differentiate between base APKs, configuration splits (e.g., for specific device architectures or languages), and dynamic feature modules.

\par\noindent\textbf{Codebase Volume Analysis.}
To quantify code volume, the \textsc{Androguard}\footnote{\url{https://github.com/androguard/androguard}} framework was applied to the Dalvik Executable (\texttt{.dex}) files. This Python-based toolkit facilitated programmatic analysis of Dalvik bytecode, enabling the extraction of quantitative metrics such as the total count of classes, methods, and fields. These indicators provided a measure of the codebase's size and complexity.

\par\noindent\textbf{Security \& Privacy Assessment.}
For a comprehensive security audit, we used the \textsc{Mobile Security Framework (MobSF)}.\footnote{\url{https://github.com/MobSF/Mobile-Security-Framework-MobSF}} MobSF automates the identification of insecure configurations, including overly permissive setups, exported components, and embedded trackers. Its vulnerability scanner cross-references API usage against known insecure patterns, checks native libraries for exploit mitigations, and performs entropy analysis to detect hard-coded secrets, providing a consolidated view of the application's security posture.

\subsection{Dynamic analysis}
\label{sec:dynamic-analysis-methods}
The dynamic analysis part of our study focuses on network traffic and privacy-relevant runtime behaviors (access to camera, microphone, location, contacts) of the applications under different scenarios (usage, permissions), presented later in this section.
\smallskip\par\noindent\textbf{Setup.}
We capture network traffic using tcpdump\footnote{\url{https://www.tcpdump.org/}} and
runtime behaviors using the SliceDroid framework~\cite{Alexopoulos2025SliceDroid}. SliceDroid performs trace \emph{slicing}, tracking inter-process communication, in order to attribute kernel events (e.g., captured by ftrace~\cite{rostedt2009finding}) to specific applications, even when these events are performed indirectly by Android Framework daemons. We choose SliceDroid as it can capture both direct and indirect resource access, possibly identifying cases where direct permission checks should have prevented access but data flow from the resource to the application takes place \emph{indirectly} via other processes. We run SliceDroid with a window size of 500 events and an overlap of 100 events.

One important challenge we faced was associating network packets captured by tcpdump to the applications under test.
To address this challenge, we extended SliceDroid to improve visibility into network operations. Specifically, we added kprobes\footnote{\url{https://docs.kernel.org/trace/kprobes.html}} on socket-related functions including \texttt{tcp\_sendmsg}, \texttt{udp\_sendmsg} and the associated \texttt{recv\_msg} functions. We matched network packets captured by tcpdump to the aforementioned kernel events based on their order, transport-layer protocol used, port number, and packet length. Finally, we use the \emph{ip-api}\footnote{\url{https://ip-api.com/}} geolocation service to get the country and organization of remote network endpoints.

\smallskip\par\noindent\textbf{Analysis Procedure.}
To systematically elicit reproducible runtime behaviors, we investigated four distinct scenarios per application: (i) \emph{foreground active with full permissions}, representing normal interactive use with all privileges granted; (ii) \emph{background passive with full permissions}, where the application remained idle in the background while retaining all permissions; (iii) \emph{foreground active with no permissions}, simulating normal usage with all available permissions explicitly revoked; and (iv) \emph{background passive with no permissions}, leaving the application idle with no privileges granted. In the foreground scenarios, each application was executed for approximately 10 seconds, during which the sole user action was composing and sending a single message. This ensured comparability across runs while reflecting a core use case. In the background scenarios, the device remained idle on the home screen for 5 minutes, with Wi-Fi enabled. Each experiment was repeated 10 times per application and per scenario to mitigate the  variability of runtime behavior. 

\smallskip\par\noindent\textbf{Significance Testing.}
To test for statistical significance of observations, we used the non-parametric Kruskal-Wallis H test~\cite{kruskal1952use} and the post-hoc Mann-Whitney U test with Bonferroni correction, since normality assumptions were violated in most cases.
\section{Experiments}
This section reports results of both static and dynamic analysis.
All experiments were conducted on the most recent stable releases available
at the time of the study, downloaded from the Google Play Store:
Messenger (523.0.0.53.109), Signal (7.54.1), and
Telegram (12.0.1), which were the three applications selected
for analysis.

\subsection{Static analysis}
Using the methods introduced in Section~\ref{sec:methods}, we focus on metrics relating to (a) application complexity, (b) attack surface, (c) permission use, and (d) static analysis warnings.

\subsubsection{Application Packaging and Complexity}
All three applications are compiled against recent Android releases (API~34 or higher), with Messenger being the only one to target API~35. 
Their minimum SDK requirements, however, diverge: Signal maintains backward compatibility down to API~21, Telegram requires API~23, and Messenger enforces API~28. 
While lower minimum SDKs broaden device compatibility, they also preserve legacy behaviors that may introduce risks \cite{android_uses_sdk}. 
 Table~\ref{tab:metadata} includes a selection of APK metadata, including the base APK size, and further distinguishes how much of the package is concentrated in the base APK versus distributed into split modules\footnote{\url{https://developer.android.com/topic/modularization}}, and how this choice affects the cumulative footprint. The reported cumulative footprint represents the total installed size, which was measured by including the necessary splits for a device with its locale set to a European country. 

\begin{table}[htbp]
\centering
\caption{Key APK metadata}
\label{tab:metadata}
\begin{tabular}{@{}lccc@{}}
\toprule
\textbf{Metric} & \textbf{Signal} & \textbf{Telegram} & \textbf{Messenger} \\
\midrule
Min SDK         & 21              & 23                & 28                 \\
Target SDK      & 34              & 34                & 35                 \\
Base APK size        & \SI{88.6}{MB}   & \SI{45.6}{MB}     & \SI{66.3}{MB}      \\
Total size with splits & \SI{115}{MB} & \SI{73.7}{MB} & \SI{68.4}{MB} \\
Number of splits & 2 & 4 & 5 \\
\bottomrule
\end{tabular}
\end{table}

As shown in the table, Signal ships only two split packages, yet its cumulative APK size surpasses both Telegram and Messenger.
This discrepancy arises from resource packaging: the compiled resource table (\texttt{resources.arsc}) aggregates localized strings, layouts, and configuration entries, and its size directly reflects how language support is delivered. Signal embeds the full set of translations within the base package, inflating the table to 27~MB—over 30\% of the base APK. By contrast, Messenger and Telegram distribute translations via split modules, keeping their base resource tables below 3~MB.

\begin{table}[htbp]
  \centering
  \caption{Code volume metrics.}
  \label{tab:code-metrics}
  \begin{tabular}{lrrr}
    \toprule
    \textbf{Metric} & \textbf{Signal} & \textbf{Telegram} & \textbf{Messenger} \\
    \midrule
    DEX files       & 7        & 5        & 11 \\
    DEX size (MB)   & 53.83    & 33.04    & 75.21 \\
    Classes         & 55{,}751 & 35{,}235 & 107{,}435 \\
    Methods         & 361{,}846 & 219{,}937 & 489{,}840 \\
    Fields          & 156{,}811 & 146{,}636 & 453{,}538 \\
    \bottomrule
  \end{tabular}
\end{table}

Regarding the size of their DEX bytecode, the three applications exhibit substantial variation (Table~\ref{tab:code-metrics}). Messenger has by far the largest footprint, with 11 DEX files totaling \SI{75.21}{MB}, containing more than \num{107000} classes and nearly half a million methods. Signal follows with 7 DEX files (\SI{53.83}{MB}), comprising about \num{55751} classes and \num{361846} methods. Telegram is the most compact, with 5 DEX files of \SI{33.04}{MB}, around \num{35235} classes and \num{219937} methods. These metrics collectively approximate the \emph{code volume} of each application: larger DEX sizes, higher class counts, and more methods correspond to a more extensive and complex compiled codebase. By this measure, Messenger’s bytecode volume is more than two times larger than Telegram’s.

The applications also diverge significantly in their native code architecture. 
Telegram adopts a monolithic approach, consolidating nearly all native 
functionality into a single shared library exceeding \SI{25}{MB}. 
In contrast, Signal takes a modular approach, distributing its cryptographic 
primitives and other functions across multiple distinct shared libraries. 

\begin{table}[htbp]
\centering
\caption{Native code summary}
\label{tab:native-summary}
\begin{tabular}{@{}lccc@{}}
\toprule
\textbf{Metric} & \textbf{Signal} & \textbf{Telegram} & \textbf{Messenger} \\
\midrule
Total native code   & \SI{24.8}{MB} & \SI{26.2}{MB} & \SI{23.3}{MB} \\
Library count       & 10            & 2             & 3+ \\
Packaging           & ARM64 split   & ARM64 split   & Compressed \\
Largest component   & \SI{7.3}{MB}  & \SI{25.3}{MB} & \SI{22}{MB}$^{*}$ \\
\bottomrule
\end{tabular}
\end{table}

\subsubsection{Attack surface metrics}
An application's attack surface is primarily defined by its exported components-entry points explicitly made accessible to other applications on the system~\cite{manadhata2011attack}. These components represent the most direct vectors for inter-app communication and potential attacks. A review of these exported components reveals significant discrepancies in the external exposure of the three apps (Table~\ref{tab:exported-components}).

Messenger presents the largest attack surface. It exposes 25 services, 15 broadcast receivers, and 7 content providers to other apps, creating channels for external interaction. Telegram also exposes a considerable number of background components, with 15 exported services and one provider. In contrast, Signal maintains a minimal external footprint, exposing only 7 services and no content providers, a design choice that prevents other applications from directly querying its data. Interestingly, the number of exported UI entry points (Activities) is comparable across the three apps.

\begin{table}[htbp]
  \centering
  \caption{Comparison of exported app components.}
  \label{tab:exported-components}
  \begin{tabular}{l r r r}
    \toprule
    \textbf{Exported Component} & \textbf{Signal} & \textbf{Telegram} & \textbf{Messenger} \\
    \midrule
    Activities & 19 & 14 & 13 \\ 
    Services & 7 & 15 & 25 \\ 
    Receivers & 4 & 3 & 15 \\ 
    Providers & 0 & 1 & 7 \\
    \bottomrule
  \end{tabular}
\end{table}

\subsubsection{Permissions}
A key aspect of an application's privacy posture is its requested permissions, which govern its access to sensitive system resources and user data. Results for all applications are shown in Table~\ref{tab:perm-aggregate}. At a high level, Telegram requests the fewest permissions overall (71), yet it also exhibits the highest count of dangerous\footnote{\url{https://developer.android.com/guide/topics/permissions/overview}} ones (25), signaling broader access to sensitive resources. Signal follows with 72 permissions, including 19 dangerous ones, representing a comparatively leaner privilege profile. Messenger, by contrast, requests the most (87) permissions in total, of which 24 are dangerous, and further stands out for requesting the most vendor-specific ``unknown’’ permissions. These ``unknown'' permissions are typically not part of the standard Android OS but are either custom permissions defined by the application itself for inter-component communication,\footnote{e.g., \texttt{org.telegram.messenger.permission.MAPS\_RECEIVE}} or vendor-specific permissions for interacting with services like the Google Services Framework.

\begin{table}[htbp]
  \centering
  \caption{Aggregate permission footprint by protection level.}
  \label{tab:perm-aggregate}
  \begin{tabular}{l r r r r}
    \hline
    \textbf{App} & \textbf{Dangerous} & \textbf{Normal} & \textbf{Unknown} & \textbf{Total} \\ \hline
    Signal    & 19 & 46 & 7  & 72 \\ 
    Telegram  & 25 & 38 & 8  & 71 \\ 
    Messenger & 24 & 46 & 17 & 87 \\ \hline
  \end{tabular}
\end{table}

Requesting \emph{dangerous} permissions—those governing access to sensitive
resources such as contacts, calendar, camera, microphone, location,
or storage—is at least to a point inevitable for modern IM clients: contact
synchronization (\texttt{READ\_CONTACTS}, \texttt{WRITE\_CONTACTS}) underpins
address-book integration, storage permissions allow media exchange, while
microphone, camera, and location access enable voice messages, video calls,
and live-location sharing.
Notably, Telegram and Messenger extend their reach with
system-level privileges such as \texttt{CALL\_PHONE},
\texttt{SYSTEM\_ALERT\_WINDOW}, and full account management---features that
power in-app calling, chat-head overlays, or multiple account workflows---but
simultaneously broaden the attack surface~\cite{Alepis2019, Fratantonio2017}.

Signal, by contrast, adopts a \emph{minimal-necessary}
stance, omitting phone-call control, overlay windows, background location,
calendar access, and package-installation rights.  
Table~\ref{tab:differential-permissions} shows notable
differences in dangerous permissions across the three apps.

\begin{table}[htbp]
  \centering
  \caption{Differential Dangerous Permissions.}
  \label{tab:differential-permissions}
  \setlength{\tabcolsep}{6pt}
  \small
  \begin{tabular}{@{}lccc@{}}
    \toprule
    \textbf{Permission}
        & \textbf{Signal}
        & \textbf{Telegram}
        & \textbf{Messenger} \\ \midrule
    \texttt{USE\_CREDENTIALS}           & \cmark & —      & — \\
    \texttt{MANAGE\_ACCOUNTS}           & —      & \cmark & \cmark \\
    \texttt{SYSTEM\_ALERT\_WINDOW}      & —      & \cmark & \cmark \\
    \texttt{CALL\_PHONE}                & —      & \cmark & \cmark \\
    \texttt{REQUEST\_INSTALL\_PACKAGES} & —      & \cmark & — \\
    \texttt{ACCESS\_BACKGROUND\_LOCATION} & —    & \cmark & — \\
    \texttt{READ\_CALL\_LOG}            & —      & \cmark & — \\
    \texttt{READ\_CALENDAR}             & —      & —      & \cmark \\
    \texttt{WRITE\_CALENDAR}            & —      & —      & \cmark \\
    \texttt{BLUETOOTH\_CONNECT}         & —      & \cmark & \cmark \\ \bottomrule
  \end{tabular}
\end{table}

\subsubsection{Security warnings and Risk}
The static analysis performed by MobSF can flag potential implementation issues (warnings). Although not all warnings reflect practically-relevant weaknesses, their number can be used as an indicator of risk~\cite{ayewah2007static}. The analysis results are summarized in Table~\ref{tab:findings-severity} and place all three applications in a Medium Risk category, with MobSF assigning a slightly lower risk score to Messenger (48/100) compared to Signal and Telegram (both 50/100). Messenger stands out with a significantly higher count of total findings, particularly in the medium-severity tier (101). In the remainder of this section, we go over some of the critical findings (high-severity), to
assess their potential practical implications.

\begin{table}[htbp]
\centering
\caption{Quantitative Summary of Static Analysis Warnings}
\label{tab:findings-severity}
\begin{tabular}{@{}lrrrr@{}}
\toprule
\textbf{Severity Level} & \textbf{Signal} & \textbf{Telegram} & \textbf{Messenger} \\ \midrule
High                    & 4               & 4                 & 9                  \\
Medium                  & 43              & 47                & 101                \\
Info                    & 3               & 4                 & 3                  \\
Secure                  & 3               & 3                 & 4                  \\
Hotspot                 & 2               & 1                 & 1                  \\ \midrule
\textbf{Total Warnings}   & \textbf{55}     & \textbf{59}       & \textbf{118}       \\ \bottomrule
\end{tabular}
\end{table}

\begin{table*}[ht]
\centering
\caption{Overview of network activity across 10 runs per scenario. Values are reported as Full~/~Restr. for the full and restricted permission scenarios. Results for TCP and UDP are reported separately in the first two sections, while the last section (ALL) sums values for both protocols. Values in bold show unique statistical discrepancies, i.e. the value for one app is significantly different from the values for the other two apps, while the values for the other two do not differ.}
\label{tab:net-activity}
\begin{tabular}{llrrrr}
\toprule
\textbf{Proto} & \textbf{Scenario} & \textbf{Metric} & \textbf{Messenger} & \textbf{Signal} & \textbf{Telegram} \\
\midrule
\multirow{6}{*}{TCP}
 & \multirow{3}{*}{\textbf{Foreground}} 
   & Peers & \textbf{4--5}~/~0 & 2~/~2--3 & 2--3~/~0--4 \\
 &  & Sent (B) & 7{,}510.5~/~0.0 & 3{,}486.0~/~3{,}535.0 & 4{,}170.0~/~12{,}552.0 \\
 &  & Recv (B) & 2{,}548.5~/~0.0 & 1{,}283.0~/~1{,}904.5 & 11{,}220.0~/~211{,}575.0 \\
\cmidrule(lr){2-6}
 & \multirow{3}{*}{\textbf{Background}} 
   & Peers & 0~/~0--5 & 0~/~0--3 & 0~/~0--3 \\
 &  & Sent (B) & 0.0~/~0.0 & 0.0~/~0.0 & 0.0~/~0.0 \\
 &  & Recv (B) & 0.0~/~0.0 & 0.0~/~0.0 & 0.0~/~0.0 \\
\midrule
\multirow{6}{*}{UDP}
 & \multirow{3}{*}{\textbf{Foreground}} 
   & Peers & \textbf{2--3}~/~\textbf{2--3} & 0~/~0--1 & 0~/~0 \\
 &  & Sent (B) & \textbf{56{,}904.0}~/~\textbf{47{,}153.5} & 0.0~/~0.0 & 0.0~/~0.0 \\
 &  & Recv (B) & \textbf{16{,}360.0}~/~\textbf{12{,}278.0} & 0.0~/~0.0 & 0.0~/~0.0 \\
\cmidrule(lr){2-6}
 & \multirow{3}{*}{\textbf{Background}} 
   & Peers & \textbf{1--3}~/~1--3 & 0--1~/~0--1 & 0--1~/~0--1 \\
 &  & Sent (B) & \textbf{1{,}548.5}~/~\textbf{1{,}694.5} & 0.0~/~0.0 & 0.0~/~0.0 \\
 &  & Recv (B) & \textbf{1{,}518.0}~/~\textbf{1{,}644.0} & 286.0~/~143.0 & 0.0~/~0.0 \\
\midrule
\multirow{6}{*}{ALL}
 & \multirow{3}{*}{\textbf{Foreground}} 
   & Peers & 2--5~/~2--3 & 0--2~/~0--3 & 0--3~/~0--4 \\
 &  & Sent (B) & 64{,}414.5~/~47{,}153.5 & 3{,}486.0~/~3{,}535.0 & 4{,}170.0~/~12{,}552.0 \\
 &  & Recv (B) & 18{,}908.5~/~12{,}278.0 & 1{,}283.0~/~1{,}904.5 & 11{,}220.0~/~211{,}575.0 \\
\cmidrule(lr){2-6}
 & \multirow{3}{*}{\textbf{Background}} 
   & Peers & 1--3~/~1--3 & 0--1~/~0--1 & 0--1~/~0--1 \\
 &  & Sent (B) & 1{,}548.5~/~1{,}694.5 & 0.0~/~0.0 & 0.0~/~0.0 \\
 &  & Recv (B) & 1{,}518.0~/~1{,}644.0 & 286.0~/~143.0 & 0.0~/~0.0 \\
\bottomrule
\end{tabular}
\end{table*}

A notable discrepancy is the handling of unencrypted network traffic. Based on the manifest analysis, only Telegram permits cleartext traffic globally by setting \texttt{usesCleartextTraffic=true}. This insecure default exposes its network communication to potential eavesdropping and tampering. Signal, in contrast, defaults to secure traffic but permits cleartext connections to a specific list of domains related to Certificate Revocation Lists (CRLs).

Messenger exhibits the most varied code-level warnings, including world-writable files and remote debugging enabled for WebViews. Such practices expand the attack surface by exposing opportunities for data tampering 
and runtime inspection ~\cite{montealegre2018security, li2017unleashing}. Most notably, its base configuration appears to trust user-installed certificates and bypass certificate pinning.
Upon further inspection, this warning proved to be a false positive. Instead of relying on the standard Android framework for its TLS connections, Messenger uses the native C++ library \texttt{Fizz} to implement the TLS-1.3 standard. This native networking stack performs its own strict, hard-coded certificate validation, effectively ignoring the more permissive settings in the Android-level XML configuration files.\footnote{\url{https://github.com/facebookincubator/fizz/blob/8c26e338ef3c85fe7815afcbb47d43f5042a8451/fizz/protocol/DefaultCertificateVerifier.h}}

An application’s attack surface and privacy posture are significantly influenced by its external dependencies. Messenger integrates multiple external SDKs for analytics and functionality, including Google Analytics and Mapbox. In contrast, Signal and Telegram declare no third-party trackers. All three applications rely on Google’s Firebase Cloud Messaging (FCM) service for push notification delivery on Android devices. 
Moreover, Samarin et al. showed that these applications do not leak sensitive information to Firebase Cloud Messaging through push notifications 
( Google’ service to send push notifications to Android devices)~\cite{samarin2024medium}.

\begin{finding} 
\textbf{Messenger} presents the largest attack surface, while requesting the most permissions and producing the highest number of static analysis warnings (\num{118}). 
\textbf{Telegram} requests the most \emph{dangerous} permissions, while producing a moderate number of warnings (\num{59}), being unique in permitting cleartext traffic by default. 
\textbf{Signal} is the leanest, with the lowest number of external dependencies, requested permissions, and warnings (\num{55}). 
\end{finding}


\subsection{Dynamic analysis}
We conduct dynamic analysis on a Nothing Phone 2a running Android 14, following the methodology of Section~\ref{sec:dynamic-analysis-methods}.

\begin{figure*}[h]
    \centering
    \includegraphics[width=0.72\textwidth]{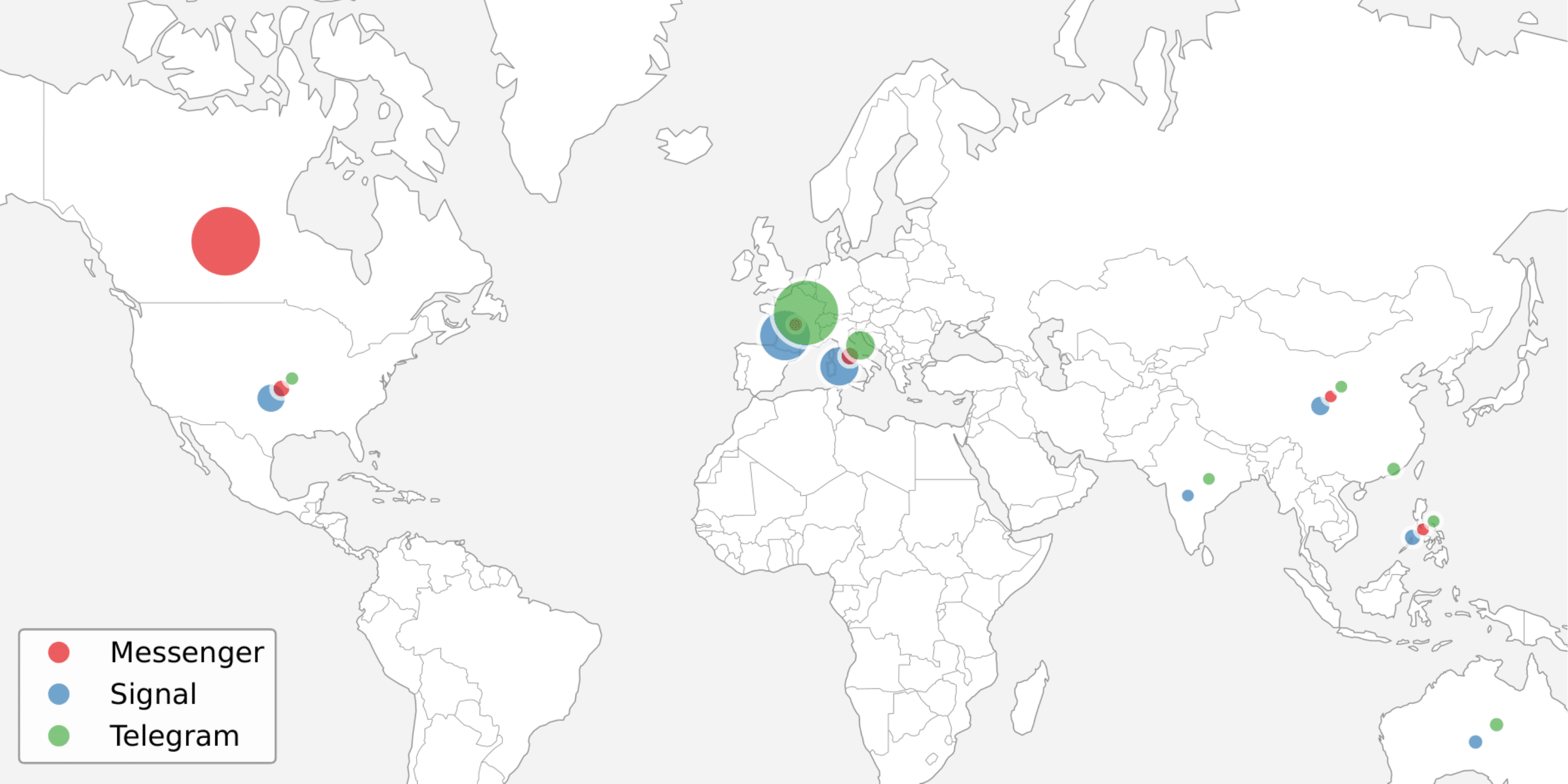}
    \caption{Geographic Traffic Distribution Map. Circle size corresponds to traffic volume.}
    \label{fig:traffic_map}
\end{figure*}

\subsubsection{Overall Network Activity}
We present a summary of our results in Table~\ref{tab:net-activity}. We report median values across ten runs for byte traffic and minimum-maximum values for the number of remote peers (distrinct IP addresses). We also separate information according to the transport-layer protocol (TCP/UDP) and test scenarios (foreground/background, full/restricted permissions). We observed the following notable discrepancies:

\smallskip\par\noindent\textbf{Overall activity.} Signal exhibited the lowest overall network activity. In foreground scenarios with permissions, Signal sent a median of almost \textasciitilde 3.5\,kB, while Messenger sent more than 64\,kB, and Telegram more than 4\,kB. Regarding inbound traffic, Signal received more  than 1\,kB, while Messenger and Telegram received 18\,kB and 4\,kb, respectively. In background scenarios, Messenger sent around 1.5\,kB of data, while Telegram and Signal did not create any traffic.

\smallskip\par\noindent\textbf{Transport-layer protocols.} Messenger relied significantly on UDP, especially in foreground scenarios with more than 80\% of its traffic being UDP. Signal and Telegram did not generate any UDP traffic in most runs (median=0). High UDP traffic generated by Messenger can be attributed to the adoption of the QUIC protocol\footnote{\url{https://engineering.fb.com/2020/10/21/networking-traffic/how-facebook-is-bringing-quic-to-billions/}}.

\smallskip\par\noindent\textbf{Effect of permissions.} Telegram is the only app that exhibited significantly higher traffic volumes while running in the foreground with no permissions (\textasciitilde 12\,kB sent, \textasciitilde 211\,kB received), compared to the same setting with full permissions (\textasciitilde 4\,kB sent, \textasciitilde 11\,kB received). Although initially counterintuitive, one likely explanation is that frequent permission denials trigger error-handling loops that involve network requests.
We leave further investigation on the exact causes of this behavior to future work.

\subsubsection{Remote Peers.}
The map of Figure~\ref{fig:traffic_map} summarizes the geographic traffic distribution of the remote endpoints contacted by the three applications, summing both inbound and outbound traffic across all scenarios. We note that our experimental location is in Europe, which likely affects results due to relay/CDN nodes. Messenger is distinct in that most traffic is exchanged with peers in North America, specifically Canada, with some additional traffic exchanged with South America and Europe. Telegram's traffic is highly concentrated in Europe with minimal additional traffic in the US, Asia, and Oceania. Signal's traffic is also concentrated mainly in Europe, albeit with significant portions in the US and Asia. Overall, Signal and Telegram maintain a more geographically distributed infrastructure compared to Messenger.

\subsubsection{Permission Compliance}
\label{sec:manual-contacts}
We verify permission compliance by inspecting the SliceDroid dashboard for accesses to restricted resources (camera, microphone, contacts, calendar) under \emph{no permissions} scenarios. Overall, the applications we tested adhered to the imposed restrictions, with no unexpected resource accesses, except for Messenger where SliceDroid flagged information flow from the contacts database.

We further investigated this anomaly by hooking and tracing all database-related methods of the system contacts process\footnote{\texttt{android.process.acore}} with the Frida dynamic instrumentation toolkit\footnote{\url{https://frida.re/}}. We consistently observed \texttt{tryWalBackgroundCheckpoint()} method calls, indicating that the observed file activity originated from the contact provider’s maintenance (Write-Ahead Logging\footnote{\url{https://sqlite.org/wal.html}}), rather than app-initiated data access. 
To further support this claim, we inspected system-level logs. \texttt{Logcat} consistently reported permission-denied requests for \texttt{READ\_\allowbreak CONTACTS}, as expected.
Overall, our manual analysis indicated that the most likely scenario is that provider-local maintenance is sometimes triggered at the time a permission request is rejected, specifically in the case of Messenger. We found no indications that the Android permission model is bypassed.
\begin{finding} 
\textbf{Messenger} sends an order of magnitude (5-10x) more data to the network in foreground scenarios compared to the other apps, with most network traffic exchanged with North America. 
\textbf{Telegram} receives almost two orders of magnitude more data (20-100x) when running in the foreground with no permissions, while most of the traffic is geographically concentrated in Europe. 
\textbf{Signal} has the smallest network footprint and the most geographically diverse infrastructure. 
\end{finding}

\section{Threats to validity} Here, we report and discuss threats to the validity of our study following established guidelines~\cite{runeson2009guidelines}.
\par\noindent\textbf{Construct validity}
In our study we employed a hybrid approach combining static and dynamic analysis tools. Although there are numerous tools available, we chose the specific ones, as they can provide the most relevant information. We make our code and data available to the community for further studies\footnote{\url{https://zenodo.org/records/17306016}}.
\par\noindent\textbf{Internal validity}
For the dynamic analysis part, patterns observed could be the result of runtime variations between the apps. We mitigate this threat by repeating experiments ten times and reporting median values instead of means. We also employ suitable statistical tests when required. Static analysis tools generally report \emph{warnings}, patterns that may point to a vulnerability or a practical problem but may also be false positives. We make this distinction clear in the paper and study specific cases in more detail.
\par\noindent\textbf{Generalization}
We examine three popular messaging apps with more than 6 billion total downloads. However, our analysis methodology can be applied to any other app and we provide our code and methodological details to the community to support further research. Furthermore, we only studied the Android implementations of the apps and further work studying implementations on other platforms would be beneficial. Finally, all network measurements were collected from one geographic location; extending the study to include multiple locations could provide a more comprehensive view of the services' infrastructure.
\section{Conclusion}
Our hybrid method combining static and dynamic analysis can identify significant privacy and security discrepancies between mobile messaging applications.
We believe that it can be applied to any messenger and also generalize to other application categories.
As might be expected from an open-source software app with an explicit
focus on privacy, Signal has the most privacy and security--friendly implementation
among the examined applications.

\medskip\par\noindent\textbf{Acknowledgments:} This project has received funding from the EU under the MSCA grant agreement No. 101108713.

\printbibliography

\appendix


\end{document}